\documentclass[aps,prb,twocolumn,showpacs]{revtex4}
\usepackage[dvips]{graphicx}

\begin{document}

\title{Nonmonotonic gap in the coexisting antiferromagnetic and
superconducting state for electron-doped cuprate superconductors}
\author{Qingshan Yuan,$^1$ Feng Yuan,$^{1,2}$ and C. S. Ting$^1$}
\address{$^1$Texas Center for Superconductivity
and Department of Physics, University of Houston, Houston, TX 77204\\
$^2$Department of Physics, Qingdao University, Qingdao 266071, China}

\begin{abstract}
We argue that the experimentally observed nonmonotonic gap in electron-doped cuprates at optimal doping is the lowest quasiparticle excitation energy in the {\it coexisting} antiferromagnetic (AF) and superconducting (SC) state. The idea is implemented by studying the coexistence of AF and SC orders with the $t$-$t'$-$t''$-$J$ model. Although the pairing gap itself is assumed to be the simplest $d$ wave which is monotonic,
we have found that the quasiparticle excitation gap in the coexisting state is nonmonotonic, with the maxima around the hot spots where the Fermi surface
is missing due to the AF gap. Within the same framework of the coexisting state the spectral function is also calculated at optimal doping. The obtained results are all consistent with experiments.
\end{abstract}

\maketitle
The pairing symmetry is a central issue for understanding the notable superconducting (SC) properties in cuprate superconductors. While for hole-doped compounds the pairing symmetry is generally accepted to be $d$ wave, it is still under debate for electron-doped ({\it e}-doped) ones.
Originally it was thought to be $s$ wave,\cite{Alff} but later suggested to be $d$ wave as in hole-doped materials, by a number of experimental measurements 
such as phase sensitive,\cite{Tsuei} angle-resolved photoemission spectroscopy (ARPES),\cite{Armitage01} penetration depth,\cite{Snezhko} etc. On the other hand, recent experimental evidence disfavoring the typical $d$ wave\cite{Blumberg,Matsui,Shan} or supporting a transition from $d$ to $s$ wave with increasing doping\cite{Biswas} was also found.
In this regard, it is of great interest to notice the measurements by Raman scattering\cite{Blumberg} and ARPES\cite{Matsui} on {\it e}-doped compounds Nd$_{2-x}$Ce$_{x}$CuO$_4$ (NCCO) at $x=0.15$
and Pr$_{1-x}$LaCe$_{x}$CuO$_4$ (PLCCO) at $x=0.11$, respectively. They are both on optimally doped samples and have reached the same conclusion,
i.e., the observed gap does not fit the simplest commonly assumed $d$-wave function $\cos k_x -\cos k_y$, but exhibits a nonmonotonic behavior with the gap maxima locating midway between the Brillouin zone (BZ) boundaries and the zone diagonals.

So far, the theoretical explanations for the so called nonmonotonic $d$-wave 
gap\cite{Yoshimura,Khodel,Watanabe,Blumberg,Matsui} are limited to extending the SC gap out of the simplest $d$ wave, e.g., by inclusion of high-order $d$-wave harmonics.

Here an alternative idea is proposed to explain the observed
nonmonotonic gap. The new idea highlights the antiferromagnetic (AF) order in the SC phase, which is motivated by the following facts.
First, in {\it e}-doped cuprates the AF order is robust to survive a broad doping range, which coexists with the SC order around 
the AF/SC phase boundary and even at optimal doping.\cite{Uefuji02,Fujita} Actually, for the
SC samples of NCCO at $x=0.15$, the N\'eel temperature $T_N$ is usually much higher than the SC transition temperature $T_c$.\cite{Uefuji02,Kang,Mang} Second, recent ARPES measurements have revealed the intriguing doping evolution of the Fermi surface (FS) in NCCO.\cite{Armitage02} 
It was found that at optimal doping the FS consists of two inequivalent pockets around $(\pi,0)$ and $(\pi/2,\pi/2)$.
This suggests the two-band modeling to study the SC properties in {\it e}-doped cuprates, as recently phenomenologically adopted by
Luo and Xiang to calculate the superfluid density.\cite{Luo} Theoretically the presence of two inequivalent FS pockets (and two bands) has been attributed to the band folding due to the AF order.\cite{Kusko,Yuan04,Yuan05,YLT}
From all of these, we argue that 
the experimentally observed gap at optimal doping is the lowest quasiparticle
excitation energy in the {\it coexisting} AF and SC state. Due to the AF gap
the FS is segmented and partly missing as seen from the ARPES,\cite{Armitage02} thus the lowest quasiparticle excitation energy is not purely the SC gap. Then, even if the SC gap itself is assumed to be the typical $d$ wave which is monotonic, the quasiparticle excitation gap may be nonmonotonic.

With the above idea in mind, in this paper we investigate the coexistence of
the AF and SC orders in {\it e}-doped cuprates, based on
the $t$-$t'$-$t''$-$J$ model. Although it is currently not reconciled whether the strong-coupling model is suitable for {\it e}-doped cuprates
(Ref.~[\onlinecite{Yuan05}] and references therein), it provides a concise
realization of the idea which is itself independent of the models. As usual, the SC pairing from the $J$ term is assumed to have the typical $d$-wave symmetry. Despite this, we have found that the 
quasi-particle excitations in the coexisting state have a nonmonotonic gap, with the maxima around the hot spots where the FS is missing. This is consistent with the experimental observation.\cite{Blumberg,Matsui} Also, the spectral function at optimal doping for NCCO is calculated for the first time within the coexisting AF and SC state. The obtained result is in agreement with the ARPES data.\cite{Armitage02}

We begin with the $t$-$t'$-$t''$-$J$ model Hamiltonian on a square lattice
\begin{eqnarray}
H & = & -t\sum_{\langle ij\rangle \sigma}(c_{i\sigma}^{\dagger}c_{j\sigma}+{\rm H.c.})
-t'\sum_{\langle ij\rangle_2\sigma}(c_{i\sigma}^{\dagger}c_{j\sigma}+{\rm H.c.}) \nonumber\\
& & -t''\sum_{\langle ij\rangle_3 \sigma}
(c_{i\sigma}^{\dagger}c_{j\sigma}+{\rm H.c.})
+J\sum_{\langle  ij\rangle }(\vec{S}_i \cdot \vec{S}_j-{1\over 4}n_i n_j)
\nonumber\\
& & -\mu_0 \sum_{i\sigma}c_{i\sigma}^{\dagger}c_{i\sigma},\label{H}
\end{eqnarray}
where 
$\langle\rangle,\ \langle\rangle_2,\ \langle\rangle_3$
represent the nearest neighbor (n.n.), second n.n., and third n.n. sites, 
respectively, and the rest of the notation is standard.
No double occupancy is allowed in the model, which implies that the Hamiltonian (\ref{H}) is already the transformed version after particle-hole transformation. Thus electron-doping already becomes hole-doping, and correspondingly one has $t<0,\ t'>0$ and $t''<0$.
Throughout the work $|t|$ is taken as the energy unit.

The slave-boson transformation is used to treat the Hamiltonian, i.e.,
$c_{i\sigma} =b_i^{\dagger} f_{i\sigma}$
with $b_i$: bosonic holon operator, $f_{i\sigma}$: fermionic spinon operator
and the constraint $b_i^{\dagger}b_i+\sum_{\sigma}f_{i\sigma}^{\dagger}f_{i\sigma}=1$
at each site. Then boson condensation is assumed, 
i.e., $\langle  b_i\rangle =
\langle  b_i^{\dagger}\rangle =\sqrt{x}$ ($x$: doping concentration) at low temperatures we are interested in. Subsequently the left spinon Hamiltonian
is decoupled with definition of the following mean-field parameters:
the uniform bond order
$\langle  f_{i\sigma}^{\dagger}f_{j\sigma}\rangle =\chi_{\sigma}=\chi,$
the AF order
$\langle f_{i\uparrow}^{\dagger}f_{i\uparrow}
-f_{i\downarrow}^{\dagger}f_{i\downarrow}\rangle/2 =(-1)^i m,$
and the spinon pairing order [or resonating-valence-bond (RVB) order]
$\langle f_{i\uparrow}f_{j\downarrow}-f_{i\downarrow}f_{j\uparrow}
\rangle=\Delta_{ij},$
where the standard $d$-wave symmetry is assumed, i.e., $\Delta_{ij}=\Delta\,(-\Delta)$ for bond $\langle ij\rangle$ along $x\,(y)$ direction. Note that the RVB order is directly associated with the SC order 
at finite doping as discussed below. We mention that the technical treatment
is similar to that used early by Inui et al.\cite{Inui} Their work, however,
does not include $t'$ and $t''$ which are essential to exhibit the electron-hole asymmetry and crucial to all the results discovered here.
In consideration of two sublattices D and E (with corresponding operators $d$ and $e$, respectively), the Hamiltonian (\ref{H}), after the above treatment, can be expanded as follows in momentum space (up to irrelevant constants)
\begin{eqnarray}
H & = & \sum_{k,\sigma} \varepsilon_k (d_{k\sigma}^{\dagger}e_{k\sigma}+{\rm H.c.})
+\sum_{k,\sigma}(\varepsilon'_k-\mu)
(d_{k\sigma}^{\dagger}d_{k\sigma}+e_{k\sigma}^{\dagger}e_{k\sigma})\nonumber\\
& & -2Jm\sum_{k,\sigma} \sigma
(d_{k\sigma}^{\dagger}d_{k\sigma}-e_{k\sigma}^{\dagger}e_{k\sigma})\nonumber\\
& & -J\sum_k \Delta_k(d_{k\uparrow}e_{-k\downarrow}+e_{k\uparrow}d_{-k\downarrow}+{\rm H.c.})\nonumber\\
& & +2NJ(\chi^2+m^2+\Delta^2/2), \label{Hk}
\end{eqnarray}
where
\begin{eqnarray*}
\varepsilon_k & = & (-2tx -J\chi)(\cos k_x +\cos k_y),\\
\varepsilon'_k & = & -4t'x\cos k_x \cos k_y - 2t''x (\cos 2k_x +\cos 2k_y),\\
\Delta_k & = & \Delta(\cos k_x -\cos k_y),
\end{eqnarray*}
$\mu$ is the renormalized chemical potential and $N$ is the total number of lattice sites. Note that the wavevector $k$ is restricted to the magnetic Brillouin zone (MBZ).

The Hamiltonian (\ref{Hk}) can be diagonalized,
giving rise to the quasiparticle energy bands $\pm E_k^{\pm}$, with
\begin{eqnarray}
E_k^{\pm} & = & \sqrt{(\xi_k^{\pm}-\mu)^2+(J\Delta_k)^2}, \label{Ek}\\
\xi_k^{\pm} & = & \varepsilon'_k\pm \sqrt{\varepsilon_k^2+4J^2m^2}.
\end{eqnarray}
And the free energy (per site) is written down
\begin{eqnarray}
F/N & = & -(2T/N)\sum_{k,\nu=+,-} \ln [2\cosh (E_k^{\nu}/2T)]\nonumber\\
& & -\mu x + 2J(\chi^2+m^2+\Delta^2/2).
\end{eqnarray}
For each given doping $x$ and temperature $T$, the order parameters
$\chi,\ m,\ \Delta$ and the chemical potential $\mu$ are determined by
minimizing the free energy.

\begin{figure}[ht]
\begin{center}
\includegraphics[width=7cm,height=8.5cm,clip]{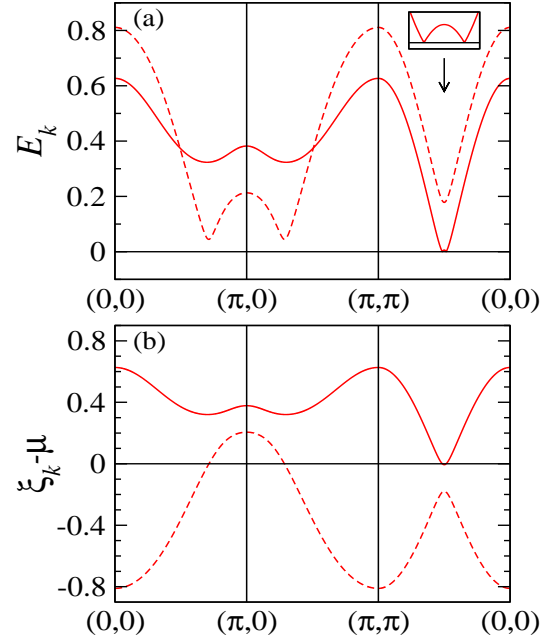}
\end{center}
\caption{Energy bands at doping $x=0.15$: (a) quasiparticle bands $E_k^{\pm}$ in the presence of both AF and SC orders (with $\Delta=0.0916,\ m=0.144,\ \chi=-0.189,\ \mu=-0.00385$) and (b) hypothetical AF bands $\xi_k^{\pm}-\mu$ in the absence of pairing (with the same $m,\ \chi$ and $\mu$ as above). In each panel the solid (dashed) line represents the band `$+(-)$'. The model parameters are: $t'=0.32,\ t''=-t'/2,\ J=0.3$ (in units of $|t|$).}
\label{Fig:Ek}
\end{figure}

\begin{figure}[ht]
\begin{center}
\includegraphics[width=5cm,height=5cm,clip]{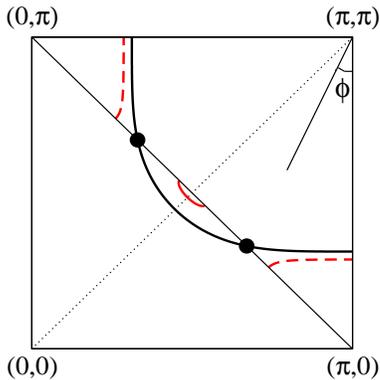}
\end{center}
\caption{Fermi surfaces in the first quadrant. The red lines give the FS for the AF bands in Fig.~\ref{Fig:Ek}(b), of which the solid line is contributed by band `$+$' and the dashed ones by `$-$'. [The FS for the `$-$' band has been moved outside of the MBZ by wavevector $(\pi,\pi)$.] For comparison, the thick black line around $(\pi,\pi)$ shows the FS for free particles (without AF order). Its crosses with the MBZ boundary are so called hot spots, as indicated by the black dots.}
\label{Fig:FS}
\end{figure}

In the following we are limited to $T\simeq 0$. And we are mainly interested in optimal doping, i.e., $x=0.15$ for NCCO.
It is expected at this doping to obtain a coexisting AF and SC state
which has nodal quasiparticle excitations.
This can be easily realized with choice of typical model parameters: $t'=0.32$, $t''=-0.16$ and $J=0.3$, leading to $\Delta=0.0916,\ m=0.144,\ \chi=-0.189,\ \mu=-0.00385$ at $x=0.15$. The quasiparticle energy bands $E_k^{\pm}$, defined in the MBZ, are extensively plotted in the original BZ
in Fig.~\ref{Fig:Ek}(a). The zero energy excitations, contributed by the `$+$' band, are present around $(\pi/2,\pi/2)$ (see the inset). This will lead to the `V-shaped' density of states at very low energy.
In order to clearly understand the formation of the quasiparticle bands, we have artificially plotted the energy bands $\xi_k^{\pm}-\mu$ with the same parameters $m$, $\chi$ and $\mu$ as above, which can be regarded as the AF bands in the absence of pairing.
Both bands are crossed by the Fermi level, around $(\pi,0)$ and $(\pi/2,\pi/2)$, respectively. The corresponding FS is plotted in Fig.~\ref{Fig:FS} in the first quadrant by the red solid and dashed lines. It consists of three pieces, 
which are separated due to the AF gap, in contrast to the continuous FS around $(\pi,\pi)$ for free particles as shown by the thick black line in Fig.~\ref{Fig:FS}.
When the pairing is considered, the RVB gap will be further opened,
leading to the quasiparticle bands in Fig.~\ref{Fig:Ek}(a).

\begin{figure}[ht]
\begin{center}
\includegraphics[width=8cm,height=6cm,clip]{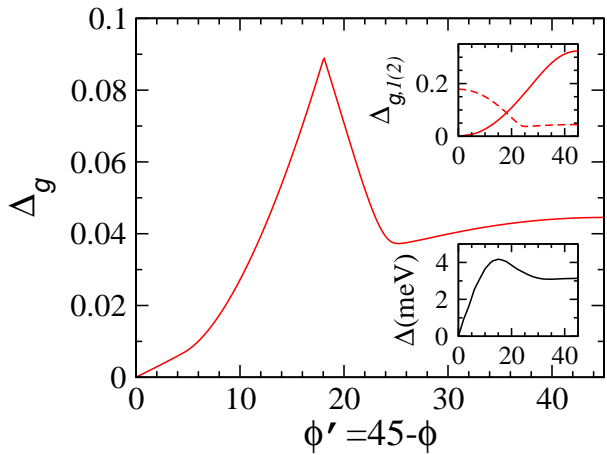}
\end{center}
\caption{Quasiparticle excitation gap $\Delta_g$ (in unit of $|t|$) as a function of $\phi'$. Upper inset: energy minima of bands $E_k^+$ (solid line) and $E_k^-$ (dashed line) in each direction $\phi'$. Lower inset: a sketch of the experimental gap for
NCCO at $x=0.15$, extracted from Ref.~[\onlinecite{Blumberg}].}
\label{Fig:Dg}
\end{figure}

From the energy bands $E_k^{\pm}$, the excitation gap $\Delta_g$ can be extracted. For a given direction as indicated by angle $\phi$ in Fig.~\ref{Fig:FS}, the momentum components $k_x$ and $k_y$ satisfy the relation $\tan \phi=(\pi-k_x)/(\pi-k_y)$. In each direction, the energy minima $\Delta_{g,1}$ and $\Delta_{g,2}$ for bands $E_k^+$ and $E_k^-$, respectively, can be found out, which are shown in the upper inset of Fig.~\ref{Fig:Dg}. The final excitation gap $\Delta_g$ in the direction $\phi$ should be the smaller one of $\Delta_{g,1}$ and $\Delta_{g,2}$, as given by the main panel of Fig.~\ref{Fig:Dg}. It is clear that $\Delta_g$ shows a nonmonotonic behavior, with the gap maximum at $\phi'=18^{\circ}$, around one of the hot spots shown in Fig.~\ref{Fig:FS}. This is qualitatively the same as the experimental results,\cite{Blumberg,Matsui} see the lower inset of Fig.~\ref{Fig:Dg} where the data from Ref.~[\onlinecite{Blumberg}] are re-plotted.
The maximum position $\phi'=18^{\circ}$ is close to $\phi'=15^{\circ}$ or $20^{\circ}$ observed in optimally doped NCCO\cite{Blumberg} or PLCCO,\cite{Matsui} respectively.

The nonmonotonic gap can be explained as follows based on the FS (red lines) in Fig.~\ref{Fig:FS}.
At $\phi=45^{\circ}\ (\phi'=0)$, i.e., in the zone diagonal direction, the FS contributed by the `$+$' band is crossed and at the same time the $d$-wave pairing gap vanishes, thus the excitation energy is zero. When $\phi$ slightly deviates from the nodal direction, the FS is still crossed and thus the excitation gap is just the pure pairing gap for the `$+$' band. Once $\phi$ decreases so that the direction points to the region where the FS is missing, the excitation gap is combined from the nonzero term $\xi_k^+ -\mu$ and the pairing gap [see Eq.~(\ref{Ek})]. Due to the extra energy $|\xi_k^+ -\mu|$ caused by the AF gap, $\Delta_g$ increases rapidly and reaches the maximum when $\Delta_{g,1}=\Delta_{g,2}$. Then $\Delta_g$ transfers from $\Delta_{g,1}$ to $\Delta_{g,2}$. Still, the current $\Delta_{g,2}$ is combined from the term
$\xi_k^- -\mu$ and the pairing gap. When $\phi$ continues to decrease, the FS contributed by the `$-$' band is crossed. And $\Delta_g$ becomes again the pure pairing gap for the `$-$' band, which is also $d$ wave,
and persists till $\phi=0\ (\phi'=45^{\circ})$.

\begin{figure}[ht]
\begin{center}
\includegraphics[width=4cm,height=4cm,clip]{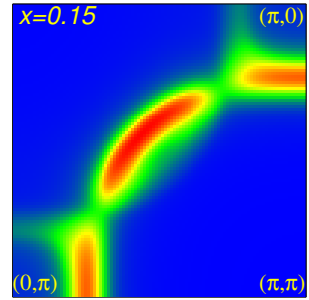}
\end{center}
\caption{Spectral intensity at doping $x=0.15$ in the coexisting AF and SC state, obtained by integrating $\bar{A}(k,\omega)$ times Fermi function over an energy interval $[-40,20]$ meV. The Lorentzian broadening is $0.03|t|$.}
\label{Fig:Akw}
\end{figure}

Within the same framework of the coexisting state, we further calculate the single particle spectral function, in order to compare with the ARPES result on NCCO at optimal doping $x=0.15$.\cite{Armitage02} The similar calculation at optimal doping was previously undertaken, but limited to the pure AF state.\cite{Kusko,Yuan04,Yuan05}
To match the experimental situation at $x=0.15$ and low temperature $T<T_c$,\cite{Armitage02} the coexisting state including the SC order is necessary. For primitive physical electrons, which are represented by operators $\bar{c}_{k\sigma}$ [equal to $c^{\dagger}_{k\sigma}$ in Hamiltonian (\ref{H})], 
the Matsubara Green's function is $\bar{G}_{\sigma}(k,\tau)=-\langle T_{\tau}\bar{c}_{k\sigma}(\tau)\bar{c}_{k\sigma}^{\dagger}(0)\rangle\simeq
-x \langle T_{\tau}f_{k\sigma}^{\dagger}(\tau)f_{k\sigma}(0)\rangle$.
The Fourier transformation of $\bar{G}$ is eventually derived, irrelevant of spin $\sigma$, as follows
\begin{eqnarray}
\bar{G}(k,i\omega_n) & = & {x\over 4}\left[(1-p_k)\left(
{1+r_k^-\over i\omega_n+E_k^-}+{1-r_k^- \over i\omega_n-E_k^-}\right)\right.\nonumber\\
& & +\left.(1+p_k)\left({1+r_k^+\over i\omega_n+E_k^+}+{1-r_k^+\over i\omega_n-E_k^+}\right)\right]
\end{eqnarray}
with the fermionic Matsubara frequencies $\omega_n$ and the coefficients
$$p_k = \varepsilon_k/\sqrt{\varepsilon_k^2+4J^2m^2},\ \ 
r_k^{\pm} = (\xi_k^{\pm} -\mu)/E_k^{\pm}.$$
The spectral function is $\bar{A}(k,\omega)=-(1/\pi){\rm Im} \bar{G}(k,i\omega_n)|_{i\omega_n\rightarrow \omega+i0^+}$.
For direct comparison with experiment, we have calculated the integration
$\int {\rm d\omega} \bar{A}(k,\omega)/(1+e^{\omega/T})$
over an energy interval $[-0.12,0.06]|t|\simeq [-40,20]$ meV and drawn its density plot in Fig.~\ref{Fig:Akw}. Strong intensity is concentrated on several separate regions, i.e., the patch around $(\pi/2,\pi/2)$, a half of the pocket around $(\pi,0)$ and equivalently that around $(0,\pi)$. All of them, when looked together, are close to a large curve around $(\pi,\pi)$. This is well consistent with the ARPES result (Fig.~3 of Ref.~[\onlinecite{Armitage02}] for $x=0.15$).

Compared to the corresponding result (not shown) calculated in the pure AF state embodied in Fig.~\ref{Fig:Ek}(b), the spectral intensity obtained here in the coexisting state is qualitatively similar, but quantitatively weaker mainly in the region around $(\pi,0)$ [and $(0,\pi)$]. This is understandable in view that the $d$-wave SC gap is actually small around $(\pi/2,\pi/2)$ and only becomes notable around $(\pi,0)$. Thus the current result has justified the part reasonability of the previous calculations at optimal doping which were based on the pure AF state.\cite{Kusko,Yuan04,Yuan05}
On the other hand, it can be imagined that if the pairing were much stronger (i.e., the order parameter $\Delta$ were much bigger) the spectral intensity would vanish around $(\pi,0)$, but remain around $(\pi/2,\pi/2)$.

A few remarks are in order. First, the simplest $d$-wave pairing is assumed here. It might be true that the pairing itself has a more complex symmetry, which is, however, not the emphasis of our idea. Second, the mean-field treatment tends to overestimate the AF order and perhaps simultaneously the SC order. Thus smaller order parameters $m$ and $\Delta$ are expected under consideration of the quantum fluctuations, leading to a reduced excitation gap as shown in Fig.~\ref{Fig:Dg}.
Finally, the spatially uniform coexistence of both AF and SC orders is implied in our calculation. Experimentally it is not conclusive\cite{Uefuji02} whether the two orders are uniformly coexistent or not. In the latter case, further study is needed to extract the excitations in the inhomogeneous phase.

In conclusion, we have studied the coexistence of the AF and SC orders 
within the $t$-$t'$-$t''$-$J$ model for {\it e}-doped cuprates.
Although the pairing gap is assumed to be the standard $d$ wave which is monotonic, we have found that the quasiparticle excitation gap in the coexisting state is nonmonotonic, with the maxima around the hot spots where the FS is missing due to the AF gap. This provides an explanation for the nonmonotonic gap observed at optimal doping. The calculated spectral intensity in the coexisting state is also consistent with ARPES result.
Our work suggests that the {\it e}-doped cuprates at optimal doping could be understood within a unified theory based on the AF and SC coexistence.

\bigskip
We would thank T. K. Lee for helpful discussion.
This work was supported by the Texas Center for Superconductivity at
the University of Houston and the Robert A. Welch Foundation under
Grant No. E-1146.

\end{document}